# Pressure-induced one-dimensional oxygen ion diffusion channel in lead-apatite

Ri He[1], Hongyu Wu[1], Xuejian Qin[1], Xuejiao Chen[1], and Zhicheng Zhong[1,2*]

[1] CAS Key Laboratory of Magnetic Materials and Devices, and Zhejiang Province Key Laboratory of Magnetic Materials and Application Technology, Ningbo Institute of Materials Technology and Engineering, Chinese Academy of Sciences, Ningbo 315201, China

[2] China Center of Materials Science and Optoelectronics Engineering, and University of Chinese Academy of Sciences, Beijing 100049, China

**Abstract**

Recently, Lee *et al.* claimed that the experimental observation of room-temperature ambient-pressure superconductivity in a Cu-doped lead-apatite ($Pb_{10-x}Cu_x(PO_4)_6O$). The study revealed the Cu doping induces a chemical pressure, resulting in a structural contraction of one-dimensional Cu-O-Cu atomic column. This unique structure promotes a one-dimensional electronic conduction channel along the *c*-axis mediated by the O atoms, which may be related to superconductivity. These O atoms occupy 1/4 of the equivalent positions along the *c*-axis and exhibit a low diffusion activation energy of 0.8 eV, indicating the possibility of diffusion between these equivalent positions. Here, using machine-learning based deep potential, we predict the pressure-induced fast diffusion of 1/4-occupied O atoms along the one-dimensional channel in $Pb_{10}(PO_4)_6O$ at 500 K, while the frameworks of Pb triangles and $PO_4$ tetrahedrons remain stable. The calculation results also indicate Cu doping can provide appropriate effective chemical pressure, indicating the one-dimensional ion diffusion channel may appear in $Pb_9Cu(PO_4)_6O$, even at ambient pressure. Our finding shows that the one-dimensional ions diffusion channel may be an important factor to affects the fabrication and electrical measurement of doped lead-apatite.

* zhong@nimte.ac.cn

# INTRODUCTION

In recent preprints [1,2], Lee *et al.* claimed that the experimental observation of room-temperature and ambient-pressure superconductivity in a Cu-doped lead-apatite ($Pb_{10-x}Cu_x(PO_4)_6O$, i.e. LK-99). The relevant experimental characterizations indicate LK-99 exhibited a sharp drop in electrical resistance to zero and diamagnetism below 400 K. This remarkable assertion swiftly garnered global attention, prompting an influx of subsequent experimental investigations [3-9] as well as theoretical researches [10-15]. These endeavors collectively aimed to replicate the superconductive sample and subsequently elucidate its intricate mechanism.

The atomic configuration of LK-99 holds paramount importance, as the original experimental papers established a connection between its superconductivity and the one-dimensional atomic cylindrical column found in its parent phase, $Pb_{10}(PO_4)_6O$ [1,2]. At room temperature, $Pb_{10}(PO_4)_6O$ has a hexagonal structure with *P6₃/m* space group [16]. The Pb and O atoms can be divided into Pb1, Pb2, O1, and O2 depending on their site symmetry. Among them, Pb2 atoms arrange into layered triangles, visually represented by blue and red markings in Fig. 1. Four O1 atoms and one P atom form tetrahedral $PO_4$ unit, and the O2 atoms align linearly along the *c*-axis. Recent density functional theory (DFT) calculations have substantiated the ability of the 2*p* orbitals of O2 atoms to establish a one-dimensional electron cylindrical conduction channel along the *c*-axis surrounded by insulating PO4 units [10,12]. In primitive unit cell, it is important to note that the O2 atom is 1/4 occupied in the four equivalent positions along the one-dimensional channel (depicted in Fig.1b). The Cu doping induces a considerable structural contraction of the one-dimensional Cu-O2-Cu column.

Although extensive theoretical researches have delved into the electronic structure of LK-99 and parent phase [10-15], the atomic configuration-related properties, especially the stability of the one-dimensional Pb2-O2-Pb2 atomic column at finite temperatures and pressures, remain enigmatic. Regrettably, structural thermodynamics and kinetics properties prove to be beyond the capacity of DFT due to the large system sizes and simulation times involved.

To solve this problem and determine the atomic configuration-related properties, we develop a machine-learning-based deep potential (DP) model for $Pb_{10}(PO_4)_6O$ using training dataset from DFT calculations. Machine-learning potential have the flexibility and nonlinearity necessary to describe complex interatomic environment from the perspective of many body using a deep neural network. Therefore, DP model is capable of large-scale atomistic dynamic simulation at finite temperatures and pressures without any empirical parameter.

Herein, we find that our DP model can describe the structure and other structure-related properties of $Pb_{10}(PO_4)_6O$ over a wide pressure range with DFT accuracy. Using the DP model, we employ deep potential molecular dynamics (DPMD) simulations to investigate the structural stability and diffusion of 1/4-occupied O2 atoms at finite temperatures and pressures. We find that the crystal structure is stable at high temperature (~ 1000 K), and it changes to disorder state above 1200 K. More importantly, we predict that the pressure of 4 GPa can induce a fast diffusion of 1/4 occupied O2 in one-dimensional channel along the *c*-axis at 500 K, while the Pb2 triangles and $PO_4$ tetrahedron framework remain stable. Our calculation results also indicate Cu doping can provide appropriate effective chemical pressure, indicating the one-dimensional ion diffusion channel may appear in LK-99, even at ambient pressure. Our finding shows that the one-dimensional atomic column is not only an electron conduction channel but also a fast ions diffusion channel.

## COMPUTATIONAL DETAILS

### A. Deep Potential of $Pb_{10}(PO_4)_6O$

Molecular dynamics (MD) simulation is an idea tool for studying thermodynamics and kinetics properties. However, a classical interatomic interaction potential for $Pb_{10}(PO_4)_6O$ was not available currently. Machine-learning based deep potential (DP) is a powerful tool for provide high-accuracy interaction potential [17,18]. The basic idea of DP is construction of deep neural network and to fit the DFT calculation data of abundant configurations. For a well-trained DP model, given any large-scale configuration, it can figure out the corresponding total energy and atomic forces at DFT-level accuracy. To obtain high accurate DP model, a large number of representative configurations are needed. Here, we adopt a concurrent learning procedure that automatically generates configurations, covering the entire potential energy surface configuration space [19]. The concurrent learning procedure contains a series of iterations, and each iteration includes three steps: (1) training the DP model from initial training dataset, (2) exploring configurations by running NPT DPMD simulations at different temperatures and pressure, and (3) labeling configurations and adding them to the training dataset, then repeating step (1) again. In the exploration step, DP model is used for MD simulations at various temperatures (10 K ~ 1800 K) and pressures (1 bar ~ 8 GPa) to extend the configuration space. We start with 1×1×2 supercell of DFT-optimized ground-state hexagonal $P6_3/m$ phase $Pb_{10}(PO_4)_6O$. After iterating this procedure 22 times, 5313 training configurations were generated. The DP model is trained using these configurations and corresponding PBE-based DFT energies, with fitting deep neural network of size (240, 240, 240). The DEEPMD-KIT code is used for training of DP model [18]. The DP compression scheme was applied in this work for accelerating the computational efficiency of the DPMD simulations [20].

**B. Density functional theory calculations**

The first-principles DFT calculations were performed using a plane-wave basis set with a cutoff energy of 520 eV as implemented in the Vienna Ab initio Simulation Package [21,22]. For bulk $Pb_{10}(PO_4)_6O$, the electron exchange-correlation potential was described using the Perdew-Burke-Ernzerhof functional [23]. The Brillouin zone was sampled with a $4 \times 4 \times 5$ Monkhorst-Pack k-point grid for the primitive hexagonal unit cell. The convergence criteria for the energy and forces were set to $10^{-6}$ eV and 0.005 eV/Å, respectively. The climbing-image nudged-elastic-band (NEB) method was used to find the minimum energy path for the diffusion of O2 atoms along the one-dimensional channel [24].

## C. Molecular dynamics simulations

The MD simulations were performed using LAMMPS code with periodic boundary conditions[25]. In exploration step, The MD simulations adopt the isobaric-isothermal (NPT) ensemble with temperature set from 10 to 1800 K and pressure set from 1 bar to 8 GPa. An isothermal-isobaric (*NPT*) ensemble with Nosé-Hoover thermostat and Parrinello-Rahman barostat are employed to control temperature and pressure, respectively [61,62]. The time step in simulations is set to 1 fs. The well-trained DP can be used to study the atomic dynamics driven by temperature and pressure via performing DPMD simulations. The DPMD simulations starting with $8 \times 4 \times 8$ supercell ($8 \times 7 \times 6$ nm$^3$) of coordinate rotation hexagonal phase containing 20992 atoms. In DPMD simulations, the equilibrium run is 50 ps, followed by a production run of 2 ns at a specified temperature and pressure to find the equilibrium state. The lattice constant, atomic distances were calculated by the average of 20000 snapshots in 2 ns of equilibrium state.

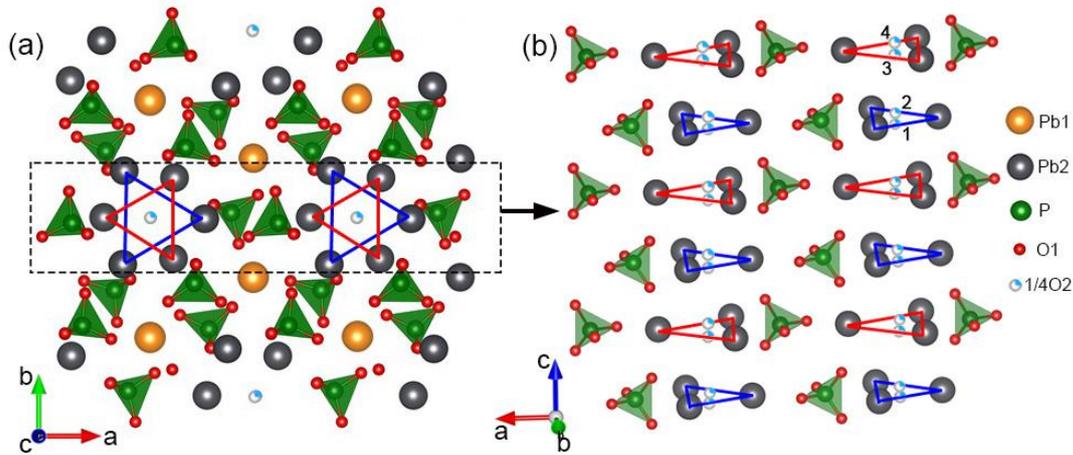

Figure 1. (a) Top view of the crystal structure of Pb$_{10}$(PO$_4$)$_6$O. The Pb2 atoms form two oppositely shaped triangles marked by blue and red lines. The O2 atoms are 1/4 occupied long the *c*-axis. The O1 atoms denote other O atoms forming tetrahedral PO4 units. (b) Side view of the crystal structure, highlighting a cylindrical column centered at 1/4 occupied O2 atoms.

## RESULTS AND DISCUSSIONS

### A. Accuracy of deep potential model

The accuracy of DP model determines the reliability of DPMD simulation. Therefore, it is first benchmarked against DFT results to confirm the accuracy of DP model. The comparison between DP and DFT calculated results for 5313 configurations is shown in Figs. 2a, b. We find a good agreement with a mean absolute error is 1.67 meV/atoms for energies and 0.068 eV/Å for forces, respectively. Table I summarizes equilibrium atomic structure and total energy optimized by DP and DFT for hexagonal phase at 0 K. All lattice constants and interatomic distance predicted by DFT and DP are almost equal, demonstrating that our DP predictions are in good agreement with DFT calculations. The equations of state of the hexagonal phase calculated by DFT and DP model are presented in Fig. 2c. The DP model reproduces well the DFT results over a wide range of lattice constants (strain of −0.07 ~ 0.06). In addition, we determined the minimum energy path of O2 atom diffusion along one-dimensional channel by using the NEB method [24]. Figure 2d shows the comparison of DFT and DP energy profiles of O2 atom diffusion pathway between 1 and 4 equivalent positions, and DP energy profiles agree with DFT results for the pathway. The calculated O2 diffusion activation energy is ~0.8 eV, indicating possibility of O2 atoms diffusion between four equivalent positions. Phonon dispersion relation is the strictest criterion for testing the accuracy of DP model. The calculated phonon dispersion relations of $1 \times 1 \times 1$ and $2 \times 2 \times 2$ hexagonal supercells by the DP model are shown in Fig. 3, which agree well with the DFT results reported by recent works [3,26]. It is interesting that the negative acoustic branches on the L-H-A ($k_z = \pi$) plane in $1 \times 1 \times 1$ supercell become positive as the unit cell is doubled as shown in Fig. 3b, this is consistent with the recent result of $Pb_{10}(PO_4)_6(OH)_2$ [3]. The systematic benchmark shows that the DP model has excellent DFT-level accuracy, and is capable of predicting a range of temperature and pressure properties of $Pb_{10}(PO_4)_6O$ from first principles.

TABLE I. Lattice constants and interatomic distances of hexagonal phase at 0 K predicted by DFT and the DP model.

|  | DFT (Å) | DP (Å) |
|---|---|---|
| Lattice $a$, $b$ | 10.030 | 10.031 |
| Lattice $c$ | 7.487 | 7.491 |
| Pb2-Pb2 in triangles 1 (marked by red in Fig.1) | 4.560 | 4.553 |
| Pb2-Pb2 in triangles 2 (marked by blue in Fig.1) | 3.943 | 3.944 |
| Pb1-Pb1 | 5.791 | 5.792 |
| Total energy (eV) | -268.5692 | -268.5686 |

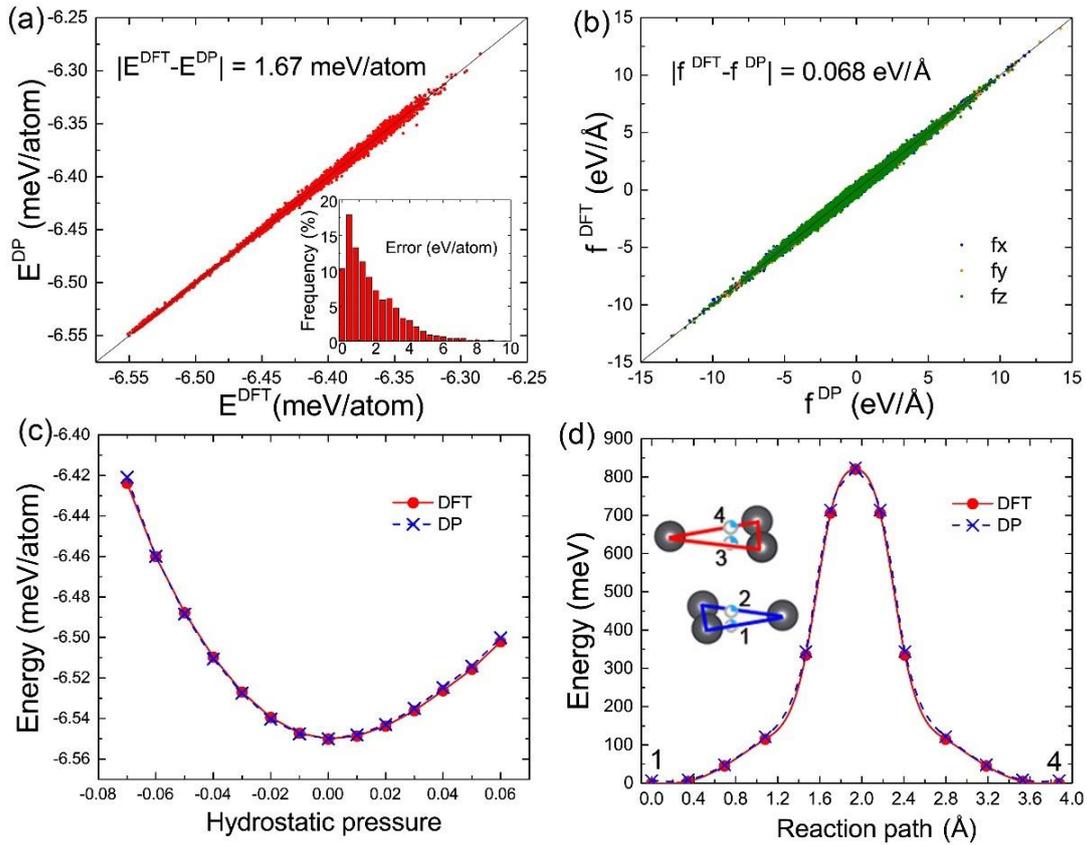

Figure 2. Benchmark test of DP against DFT results. Comparison of (a) energies and atomic (b) forces calculated using the DP and DFT for 5313 configurations in training dataset. DP and DFT energy variation for different (c) hydrostatic pressure and the (d) O2 migrations along translation pathway with minimum barrier.

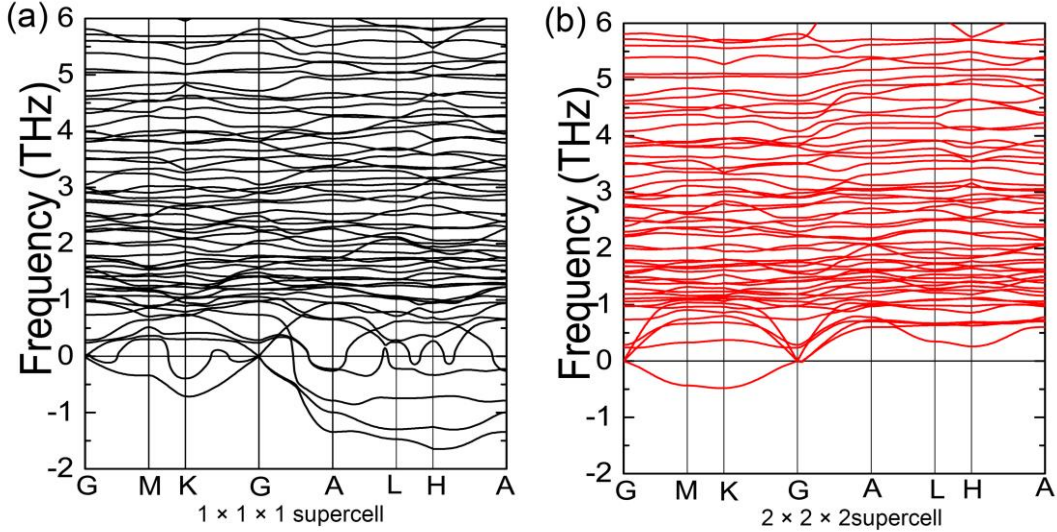

Figure 3. Phonon dispersion relations for relaxed hexagonal phase using (a) 1 × 1 × 1 and (b) 2 × 2 × 2 supercell by DP model. The negative acoustic branches on the L-H-A plane in 1 × 1 × 1 supercell become positive as the cell is doubled.

## B. Thermal stability of hexagonal $Pb_{10}(PO_4)_6O$

We perform *NPT* DPMD simulations of 20992-atom supercell with increasing temperature from 0 K to 1800 K to explore the thermal stability of $Pb_{10}(PO_4)_6O$. The lattice constant of *c*-axis as a function of temperature is shown in Fig. 4a. With the increase of temperature to 1000 K, it can be observed that the lattice constant of c-axis increases slightly. The corresponding snapshot of atomic structures (inset of Fig. 4a) and the partial radial distribution functions (RDFs) of Pb-Pb and Pb-P in Fig. 4b clearly show that the hexagonal $Pb_{10}(PO_4)_6O$ maintains perfect crystalline structure at 1000 K. Increasing the temperature beyond 1200 K, the Pb-Pb and Pb-P RDFs peaks vanish at distances larger than 8 Å (see Fig. 4b). This substantial modification indicates that $Pb_{10}(PO_4)_6O$ melts into liquid. Accompanying solid-liquid transition is a sudden increase of lattice constant from 7.7 to 8.0 Å. The snapshots of corresponding atomic structures in inset also clearly show that a disordered structure forms above 1200 K. Therefore, the DP predicted melting temperature of $Pb_{10}(PO_4)_6O$ is ~1200 K.

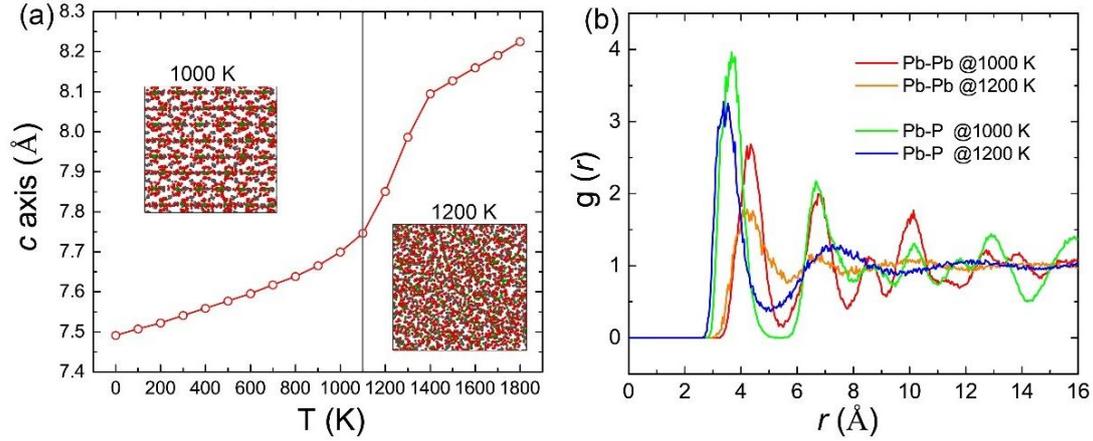

Figure 4. Crystal structural properties of hexagonal $Pb_{10}(PO_4)_6O$ on heating. (a) Variation of lattice parameter of c-axis at temperatures from 0 to 1800 K. The insets show the partial snapshot of atomic structures at 1000 K and 1200 K. (b) Radial distribution function $g_{ij}(r)$ of the atomic structures at 1000 K and 1200 K.

## C. High Oxygen ion diffusivity in one-dimensional channel.

O2 atom is 1/4 occupied in the four equivalent positions along the *c*-axis and its diffusion activation energy is ~0.82 eV. It supports a diffusion of oxygen ions between these equivalent positions along the one-dimensional channel. DPMD simulations allow a deeper understanding of oxygen ion diffusion. The ion diffusivity can be determined from the evolution of the mean square displacement (MSD) as a function of simulation time at finite temperature and pressure. Figure 5a, b compares the MSD curves of O2 in the *a-b* plane and along the *c*-axis under ambient pressure and 4GPa at 500 K. At ambient pressure, the average MSD of O1 and O2 atoms are the constants with time and the magnitude is small. It indicates that the vibration of oxygen atoms is confined in local area and without diffusion. Specifically, for O2 atom, the MSD along *c* axis significantly larger than in *a-b* plane, indicating the anisotropy of vibration and the amplitude is large along *c* axis. Figure 5c presents the typical trajectories of marked O1 and O2 (marked by yellow) atoms. It confirms again that all the oxygen atoms vibrate at local area and do not diffuse to the neighboring positions at ambient pressure.

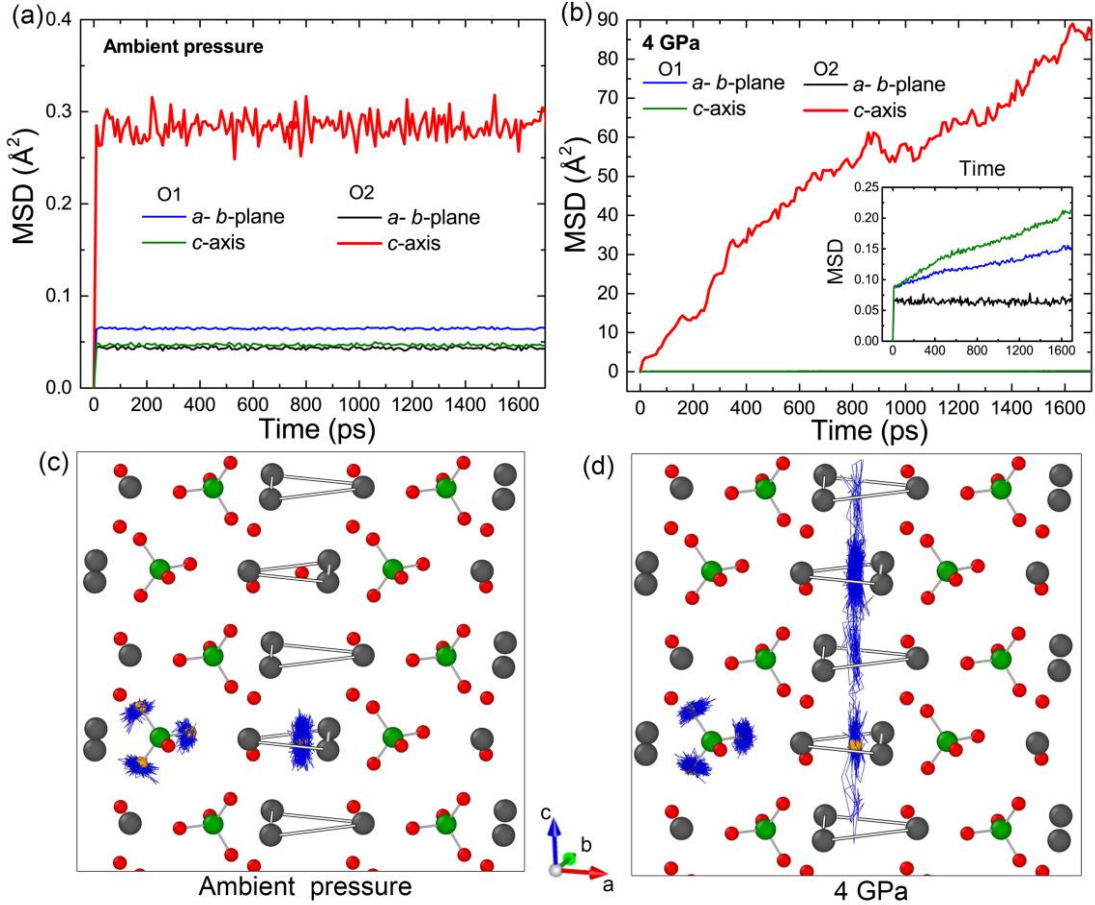

Figure 5. Mean square displacement (MSD) of O1 and O2 atoms in the *a-b* plane and along the *c*-axis under (a) ambient pressure and (b) 4GPa. Trajectories of O1 and O2 atoms marked by yellow in hexagonal $Pb_{10}(PO_4)_6O$ under (c) ambient pressure and (d) 4GPa. All the DPMD simulations were performed at 500 K.

Interestingly, upon application of 4GPa pressure at same temperature (500 K), the MSD of O2 along *c*-axis is increased with the time linearly, while in *a-b* plane the MSD remains constant that does not change with time (see Fig. 5b). It indicates that the O2 atoms diffuse along *c*-axis. The corresponding migration trajectories of O2 (marked by yellow) is shown in Fig. 5d. The O2 atom diffuses away from the initial equilibrium position to neighboring equilibrium position, then hopping between several positions along *c*-axis. In this process, these O2 atoms will exchange their position, but they do not diffuse in *a-b* plane, which means all O2 atoms were confined in one-dimensional channel. O2 atoms diffuse in the one-dimensional channel with a diffusion coefficient of $7.3 \times 10^{-6}$ $cm^2s^{-1}$, which is almost same order of magnitude as liquid [27]. Meanwhile, the O1 atoms formed tetrahedral PO$_4$ units and Pb-Pb triangles remain quite rigid under pressure. The fast O2 diffusion may be attributed to the pressure lowers the diffusion activation energy along *c*-axis. Figure 6 presents the migration trajectories of all O atoms in a slice of hexagonal stricture supercell at 500 K and 4 GPa, and it clearly conforms the one-dimensional channels of O2 diffusion. These results indicate that the pressure can induce a 1/4-occupied O2 atoms diffusion in one-dimensional channel just like liquid, while the framework is still rigid and it will maintain crystal.

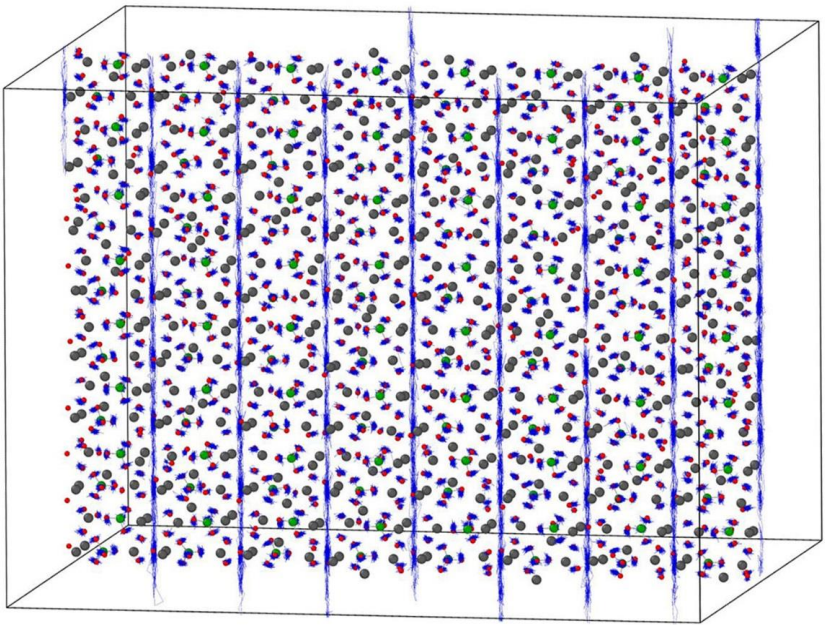

Figure 6. DPMD calculated all oxygen (O1 and O2) atoms trajectories in slice of hexagonal supercell at 500 K and 4 GPa. It clearly indicates that the one-dimensional diffusion channels of O2 diffusion.

The experimental papers reported the atomic structure is entirely affected by the stress and strain created by Cu doping [1,2]. Therefore, an appropriate structural contraction was generated in LK-99 at ambient pressure. The structural change mainly affects the contraction of the Pb atoms formed triangle. Therefore, we identify Pb-Pb distances as a descriptor for lattice contraction. To evaluate the effective chemical pressure of Cu doping, we perform DFT calculation with different hydrostatic pressures. Figure 7 shows the variation of the calculated Pb-Pb distance under different hydrostatic pressures. One can observe a linearly decrease of Pb-Pb distance with increasing pressure. We found Cu doping shrinks the Pb-Pb distance to 4.36 Å, which is equivalent to hydrostatic pressure of 2.45 GPa. It indicates the one-dimensional diffusion channel was considered to appear in Cu-doped $Pb_{10}(PO_4)_6O$, even at ambient pressure.

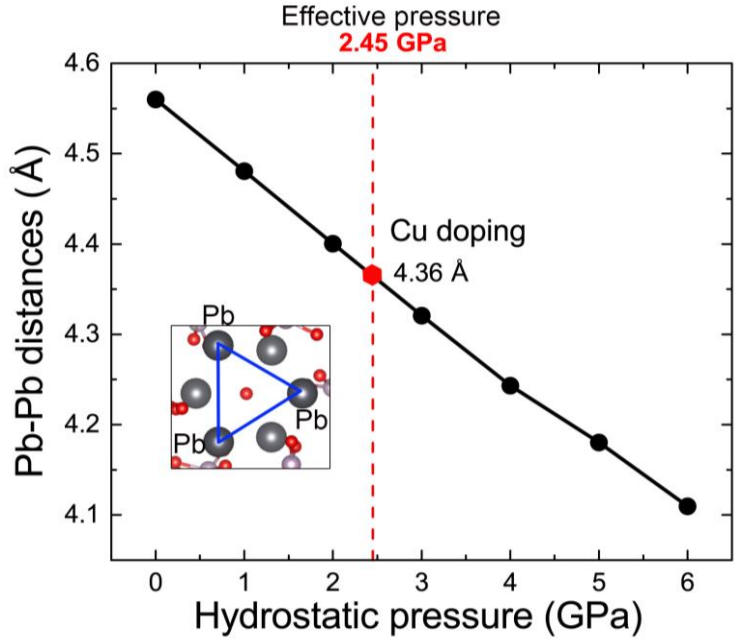

Figure 7. Pb-Pb distances for diffident hydrostatic pressures and the effect of Cu doping on the structure is equivalent to the pressure of ~2.45 GPa.

## CONCLUSION

In summary, we developed a machine-learning based DP model to describe energetic and dynamic properties of $Pb_{10}Cu(PO_4)_6O$ (parent phase of LK-99), which has DFT-level accuracy at the PBE level. Then we used the DP model to simulate temperature-driven solid-liquid phase transition, and we found that the hexagonal phase change to disorder state at 1200 K. More importantly, DPMD simulations predict that the pressure-induced strongly anisotropic diffusion of 1/4-occupied O2 atoms, resulting in the diffusion of O2 atoms in one-dimensional along the *c*-axis. However, the Pb2 triangles and $PO_4$ tetrahedron framework remain rigid under pressure. According to DFT calculations, Cu doping generates the contraction of Pb-Pb triangular structure under ambient pressure, which is equivalent to hydrostatic pressure of 2.45 GPa. It indicates the pressure and dopant would be useful for anisotropic transportation of ions at the nanometer level in specific material. Pressure-induced anisotropic transportation opens the possibilities for design of innovative microelectronic device, such as memristor electrochemical ionic synapse.

All the input files, final training datasets, and DP model files to reproduce the results contained in this paper are available in AIS Square website (https://www.aissquare.com/).

## ACKNOWLEDGMENTS

This work was supported by the National Key R&D Program of China (Grants No. 2021YFA0718900, No. 2022YFA1403000, No. 2021YFE0194200), the Key Research Program of Frontier Sciences of CAS (Grant No. ZDBS-LY-SLH008), the National Nature Science Foundation of China (Grants No. 11974365, No. 12204496, No 12161141015), the K.C. Wong Education Foundation (Grant No. GJTD-2020-11), and the Science Center of the National Science Foundation of China (Grant No. 52088101).

Journal of Computational Physics **117**, 1 (1995).

[26] J. Shen, D. Gaines, II, S. Shahabfar, Z. Li, D. Kang, S. Griesemer, A. Salgado-Casanova, T.-c. Liu, C.-T. Chou, Y. Xia *et al.*, Phase Stability of Lead Phosphate Apatite $Pb_{10-x}Cu_x(PO_4)_6O$, $Pb_{10-x}Cu_x(PO_4)_6O(OH)_2$, and $Pb_8Cu_2(PO_4)_6$, (2023), p. arXiv:2308.07941.

[27] D. F. Othmer and M. S. Thakar, Correlating diffusion coefficient in liquids, Industrial & Engineering Chemistry **45**, 589 (1953).